\documentclass[11pt,english]{article}
\usepackage[latin1]{inputenc}
\usepackage[english]{babel}
\usepackage{booktabs}
\usepackage{amsmath,amssymb,amsthm,amsfonts}
\usepackage{enumerate}
\usepackage{tikz}
\usepackage[authoryear,round]{natbib}
\usepackage{hyperref}
\usepackage[a4paper,body={16cm,23.7cm}]{geometry}
\usetikzlibrary{positioning,decorations.pathreplacing}

\hypersetup{colorlinks=false,linkcolor=cyan,citecolor=cyan}

\newtheorem{theorem}{Theorem}

\newtheorem{proposition}{Proposition}
\newtheorem{remark}{Remark}
\newtheorem{example}{Example}

\DeclareMathOperator{\DM}{DM}

\DeclareMathOperator{\NH}{NH}

\begin{document}

\title{\textbf{ORDER PRESERVATION WITH DUMMIES IN THE MUSEUM PASS PROBLEM}}
	\author{Ricardo Mart\'inez\thanks{\textbf{Corresponding author}. Departmento de Teor\'ia e Historia Econ\'omica. Universidad de Granada, Facultad Ciencias Econ\'omicas y Empresariales, 18011 Granada, Spain.} \\ Universidad de Granada
	\and Joaqu\'in S\'anchez-Soriano\thanks{Centro de Investigaci\'on Operativa (CIO), Universidad Miguel Hern\'andez de Elche, 03202 Elche, Spain.} \\ Universidad Miguel Hern\'andez de Elche}
	\date{}
	\maketitle

	\begin{abstract}
		We study the problem of sharing the revenue obtained by selling museum passes from the axiomatic perspective. In this setting, we propose replacing the usual \emph{dummy} axiom with a milder requirement: \emph{order preservation with dummies}. This new axiom formalizes the philosophical idea that even null agents/museums may have the right to receive a minimum allocation in a sharing situation. By replacing \emph{dummy} with \emph{order preservation with dummies}, we characterize several families of rules, which are convex combinations of the uniform and Shapley approaches. Our findings generalize several existing results in the literature. Also, we consider a domain of problems that is richer than the domain proposed by \cite{Ginsburgh03} in their seminal paper on the museum pass problem.
	\end{abstract}



	\newpage


	\section{Introduction}

	\cite{Ginsburgh03} introduced the so-called museum pass problem, whose goal is to share the revenue obtained by selling museum passes that allow entrance into a set of museums participating in the program. This particular setting can be extended to many real-life situations where selling products in a package is more profitable than selling them independently (see \cite{Adams76}). Recently, new media and entertainment platforms have achieved a significant success based on the same idea. For example, Netflix provides access to a collection of movies and series at a fixed price. Customers can decide on the contents of the catalog that they want to watch. The sharing problem arises when Netflix must distribute the obtained revenue among the content producers, taking into account the views.

	A museum pass problem is described by four elements: the set of museums that participate in the program, the set of pass holders, the pass price, and the entrance matrix that specifies the museums each pass holder has visited. A \emph{rule} is a method to distribute the revenue among the museums. Several authors have addressed this problem from different perspectives. \cite{Ginsburgh03} recommended the Shapley value as a natural method for distributing revenue among the participants.\footnote{Although alternatives have been analyzed (see, for instance, \cite{Bergantinos16} \cite{Wang11}, or \cite{Beal10}), the Shapley value has emerged as the convenient sharing method (see \cite{Roth88} and \cite{Algaba19b}).} Using a different approach, \cite{Estevez-Fernandez12} and \cite{CasasMendez11} interpreted the museum pass problem as a bankruptcy problem and endorsed some rules from this literature. We propose two novelties with respect to the previous works: firstly, we replace one of the necessary axioms for the characterization of the Shapley value; and secondly, we expand the domain of problems.

	\cite{Bergantinos15} proved that the Shapley value is the unique rule that satisfies three standard axioms: \emph{equal treatment of equals}, \emph{dummy}, and \emph{revenue additivity}. \emph{Dummy} states that when a museum has not been visited by any pass holder and, therefore, has not contributed to the revenue, its allotment must be zero. However, from a social welfare perspective, this straightforward implication is not so evident and desirable. This requirement asserts that individuals who have not contributed to society do not have to right to receive something from society. In other words, they are excluded. Minimum wages, universal healthcare systems, and, in general, the welfare state are counterexamples to this idea. Thus, as an alternative to \emph{dummy}, we propose \emph{order preservation with dummies}. This axiom assumes that even a \emph{dummy} museum may receive a positive allocation (obviously at the cost of decreasing the allotments of the other museums) as long as it is less than the allocation of any other \emph{non-dummy} museum. We characterize the family of rules that satisfy \emph{equal treatment of equals}, \emph{order preservation with dummies}, and \emph{revenue additivity}. Interestingly, this family of rules includes the Shapley value, the uniform distribution, and compromises between of those. We also consider the principle of \emph{$\tau$-order preservation with dummies}, which places an upper bound on the social solidarity and, therefore, the transfers from the non-dummy museums to the dummy museums.

	In the standard version of the model, it is assumed that each museum must be visited by at least one pass holder. However, this does not always need to be the case. It is not unrealistic to suppose that among the thousands of passes sold, certain visitors may not go to any museums. Similarly, it is not unrealistic to consider that among the millions of Netflix subscribers, some may not watch any content during the billing period. In this paper, we extend the problem domain to encompass these situations. Although this may appear to be a minor change, it has far-reaching implications.\footnote{A \emph{reduced domain} is the domain of problems in which each pass holder must visit at least one museum. An \emph{enlarged domain} is the domain of problems without that restriction.} We show that, on the enlarged domain, there is no rule that satisfies \emph{equal treatment of equals}, \emph{dummy}, and \emph{revenue additivity}. This is the counterpart to the characterization in \cite{Bergantinos15} for the reduced domain, on which those three same axioms characterize the Shapley value. An immediate question arises regarding the possibility of extending the Shapley value to the enlarged domain while keeping most of its essence. The answer to this question is the \emph{equal attribution rule}. On this domain, we also identify the family of rules fulfilling \emph{equal treatment of equals}, \emph{order preservation with dummies}, and \emph{revenue additivity}.

The museum pass problem is closely related to attribution problems (see \cite{Algaba19}), which are relevant in many contexts. These include the carpool problems (i.e., assigning drivers to subsets of participants who commute regularly) analyzed in \cite{Naor05}, analysis of the relevance of genes (\cite{Moretti07}), ranking wines (\cite{Ginsburgh12}), the problem of allocating publication credit among co-authors (\cite{Karpov14}), ranking of languages in the European Union before and after Brexit (\cite{Ginsburgh17}), distributing revenues from TV broadcasting (\cite{LopezNavarrete19} and \cite{Bergantinos22b}), ranking candidates when approval voting is used and the principle of \emph{one-person-one-vote} is required (\cite{Dehez20}), analysis of lethality relevance in co-morbidity (\cite{Martinez21}), or the construction of indicators to evaluate the impact of COVID-19 (\cite{Martinez22d}). Each of these problems is of great interest in its field. Therefore, it is important to analyze these problems from a theoretical perspective and to investigate possible applications considering broader domains.

	The rest of the paper is organized as follows: in Section 2, we introduce the model; in Section 3, we present several rules; Section 4 is devoted to the axioms we analyze in this paper; in Section 5, we show our characterization results; and in Section 6, we conclude the paper with some final remarks.


	\section{Benchmark model}\label{model}

	Let $\mathbb{M}$ represent the set of all potential museums and let $\mathcal{M}$ be the set of all finite (non-empty) subsets of $\mathbb{M}$. Now, let $\mathbb{N}$ represent the set of all potential buyers of a museum pass. Let $\mathcal{N}$ be the set of all finite (non-empty) subsets of $\mathbb{N}$. A (museum) \textbf{problem} is a 4-tuple $(M,N,\pi,E)$, where
	\begin{itemize}
		\item $M=\{1,\ldots,m\} \in \mathcal{M}$ is the set of \textbf{museums}.
		\item $N=\{1,\ldots,n\} \in \mathcal{N}$ is the set of \textbf{pass holders}.
		\item $\pi \in \mathbb{R}_{++}$ is the \textbf{pass price}.
		\item $E \in \{0,1\}^{n \times m}$ is the \textbf{entrance matrix} (of $n$ rows and $m$ columns), where
		$$
		E_{ai}=
		\begin{cases}
		1 & \text{if $a$ has visited museum $i$} \\
		0 & \text{otherwise}
		\end{cases}
		$$
	\end{itemize}
	We denote by $e_i=\sum_{a \in N} E_{ai}$ the number of pass holders who visited museum $i\in M$, and by $e^a=\sum_{i \in M} E_{ai}$ the number of museums visited by pass holder $a \in N$.

	The domain of all possible problems $(M,N,\pi,E)$ is $\mathcal{D}$. \cite{Ginsburgh03} and \cite{Bergantinos15} assumed that any pass holder visits, at least, one museum. We denote by $\overline{\mathcal{D}} \subset \mathcal{D}$ such a subdomain of problems:
	$$
	\overline{\mathcal{D}} = \left\{ (M,N,\pi,E) \in \mathcal{D}: e^a >0 \; \; \forall a \in N \right\}
	$$

	We refer to $\mathcal{D}$ and $\overline{\mathcal{D}}$ as the \textbf{enlarged domain} and \textbf{reduced domain}, respectively.

	Given a problem, a \textbf{null (pass) holder} is a pass holder that has not visited any museum. That is to say, $a \in N$ is null if $e^a=0$. Analogously, we say that a museum $i \in M$ is \textbf{dummy} if it has not been visited by any pass holder, that is, $e_i=0$. We denote by $\NH$ and $\DM$ the set of all null holders and dummy museums, respectively:
	$$
	\NH(M,N,\pi,E)=\left\{ a \in N: e^a=0 \right\} \quad \text{ and } \quad \DM(M,N,\pi,E)=\left\{ i \in M: e_i=0 \right\}
	$$

	A \textbf{rule} is a method to distribute the revenue obtained by selling the passes, $n\pi$, among the museums. Formally, it is a mapping $R: \mathcal{D} \longrightarrow \mathbb{R}^m_+$ such that
	$$
	\sum_{i \in M} R_i(M,N,\pi,E)=n\pi,
	$$
	where the vector $R(M,N,\pi,E) \in \mathbb{R}^m_+$ indicates the allocation of each museum.


	\section{Allocation rules}\label{MPP-secrules}

	In this section, we present some examples of rules. Some apply to the domain $\mathcal{D}$, while others are only defined on the reduced domain $\overline{\mathcal{D}}$. The first rule is straightforward. It distributes the revenue uniformly among the museums.

	\textbf{Uniform rule}. For each $(M,N,\pi,E) \in \mathcal{D}$ and each $i \in M$,
	$$
	R^U_i (M,N,\pi,E) = \frac{n \pi}{m}
	$$

	The next rule is also very natural since it divides the revenue proportionally to the visitors each museum has. Eventually, it may happen that all pass holders are null. In such an extreme and unlikely case, the rule applies an equal split.

	\textbf{Proportional rule}. For each $(M,N,\pi,E) \in \mathcal{D}$ and each $i \in M$,
	$$
	R^P_i(M,N,\pi,E) = \frac{e_i}{\sum_{j=1}^m e_j} n \pi,
	$$
	and $R^P_i(M,N,\pi,E)=\frac{n \pi}{m}$ if $E=0$.

	The third rule was proposed by \cite{Ginsburgh03}. It allocates the price of each pass equally among the museums visited by the pass holder.\footnote{\cite{Ginsburgh03} show that this rule coincides with the Shapley value (\cite{Shapley53}) of the TU-game where, for each $S \subseteq M$, $v(S) = \pi \cdot \left| \left\{ a \in N: E_{ai}=1 \; \text{ for some } i \in S \right\} \right|$.}

	\textbf{Shapley rule}. For each $(M,N,\pi,E) \in \overline{\mathcal{D}}$ and each $i \in M$,
	$$
	R^{Sh}_i(M,N,\pi,E) = \sum_{a=1}^n \frac{E_{ai}}{e^a} \pi
	$$
	Notice that the Shapley rule is only defined on the reduced domain $\overline{\mathcal{D}}$. As illustrated below, there exist several alternatives to extend this rule to the domain $\mathcal{D}$ and encompass problems with null pass holders.

	\begin{example}\label{MPP_Ex1}
		Let us consider the museum pass problem with museums $M=\{1,2,3\}$, pass holders $N=\{1,2,3,4,5\}$, pass price $\pi=1$, and an entrance matrix given by
		$$
		E=
		\left(
		\begin{array}{ccc}
		1 & 0 & 0 \\
		1 & 1 & 0 \\
		0 & 1 & 0 \\
		0 & 1 & 0 \\
		0 & 0 & 0 \\
		\end{array}
		\right)
		$$
		In this problem, the revenue to distribute is $5$. Its distributions according to the uniform and proportional rules are $\left( \frac{5}{3}, \frac{5}{3}, \frac{5}{3} \right)$ and $\left( \frac{10}{5}, \frac{15}{5}, 0 \right)$, respectively. The application of the Shapley rule, however, is not so obvious. The fifth holder is null ($e^5=0$) and, hence, $(M,N,\pi,E) \notin \overline{\mathcal{D}}$. Nevertheless, the revenue generated by this pass holder must still be divided among the museums. Several alternatives emerge as natural extensions of the Shapley rule, some of which are presented below.
	\end{example}

	If a pass holder $a \in N$ has visited at least one museum, then the next rule equally splits the price of their pass, $\pi$, among them. If, instead, $a$ is null, then $\pi$ is equally split among all museums. Formally,

	\textbf{Equal attribution rule}. For each $(M,N,\pi,E) \in \mathcal{D}$ and each $i \in M$,
	$$
	R^{EA}_i(M,N,\pi,E) = \sum_{a=1}^n \rho_i^a(M,E) \pi,
	$$
	where
	$$
	\rho^a_i(M,E)=
	\begin{cases}
	\dfrac{E_{ai}}{e^a} & \text{if } e^a >0 \\[0.3cm]
	\dfrac{1}{m} & \text{otherwise}

\end{cases}
$$

After applying the equal attribution rule to Example \ref{MPP_Ex1}, each museum receives $\frac{1}{3}$ from the pass paid by the fifth pass holder. However, one could argue that Museum 3 should be excluded from the assignment since none of the other pass holders visit it. In other words, the contribution of the null pass holders to the overall revenue must be split equally among all non-dummy museums, only. This is what the following rule does,

\textbf{Conditional equal attribution rule}. For each $(M,N,\pi,E) \in \mathcal{D}$ and each $i \in M$,
$$
R^{CEA}_i(M,N,\pi,E) = \sum_{a=1}^n \rho^a_i(M,N,\pi,E) \pi,
$$
where
$$
\rho^a_i(M,E)=
\begin{cases}
\dfrac{E_{ai}}{e^a} & \text{if } e^a >0 \\[0.3cm]
\dfrac{1}{|N \setminus \DM(M,N,\pi,E)|} & \text{if } e^a =0 \text{ and } i \notin \DM(M,N,\pi,E) \\[0.3cm]
0 & \text{otherwise}
\end{cases}
$$
and $R^{CEA}_i(M,N,\pi,E)=\frac{n \pi}{m}$ if $E=0$.

Finally, the next rule allocates the pass of the null pass holders proportionally to the visits on the non-null pass holders. Formally,

\textbf{Proportional attribution rule}. For each $(M,N,\pi,E) \in \mathcal{D}$ and each $i \in M$,
$$
R^{PA}_i(M,N,\pi,E) = \sum_{a=1}^n \rho_i^a(M,E) \pi,
$$
where
$$
\rho^a_i(M,E)=
\begin{cases}
\dfrac{E_{ai}}{e^a} & \text{if } e^a >0 \\[0.3cm]
\dfrac{e_i}{\sum_{j=1}^m e_j} & \text{otherwise}
\end{cases}
$$
and $R^{PA}_i(M,N,\pi,E)=\frac{n \pi}{m}$ if $E=0$.

It is obvious that the three previous rules are extensions to $\mathcal{D}$ of the Shapley rule, and they coincide with it when $(M,N,\pi,E) \in \overline{\mathcal{D}}$. Which one is more suitable? Section 5.2 is devoted to answer this question.


\section{Properties}

We now introduce several axioms that are suitable for this framework.

The first axiom is a standard principle of impartiality. It simply requires that if two museums have the same visitors, then they must receive equal awards.

\textbf{Equal treatment of equals}. For each $(M,N,\pi,E) \in \mathcal{D}$ and each $\{i,j\} \subseteq M$, if $E_{ai}=E_{aj}$ $\forall a \in N$, then
$$
R_i(M,N,\pi,E)=R_j(M,N,\pi,E)
$$

The next property says the allocation is independent of the timing. Imagine the following two alternatives. One, we distribute the revenue every semester, being $N$ and $N'$ the two disjoint groups of pass holders in each period. And two, we solve the problem once a year, considering the revenue generated by the whole group of pass holders, $N \cup N'$. In both cases the allocation must be the same. This axioms was introduced by \cite{Ginsburgh01} for this setting.\footnote{Relating the outcome of a group of agents with the outcome of subgroups is a usual approach in the literature. \emph{Revenue additivity} is akin to those namesakes used in the classical literature of income inequality measurement (e.g., \cite{Bourguignon79}), poverty measurement (e.g., \cite{Foster84}), income mobility measurement (e.g., \cite{Fields96}), or voting (e.g., \cite{Smith73})}

\textbf{Revenue additivity}. For each $(M,N,\pi,E), (M,N',\pi,E') \in \mathcal{D}$ such that $N \cap N' = \emptyset$, it holds that
$$
R(M,N \cup N',\pi,E \oplus E') = R(M,N,\pi,E) + R(M,N',\pi,E'),
$$
where $E \oplus E' \in \{0,1\}^{(n+n') \times m}$ is the matrix resulting from stacking $E$ above $E'$ (by rows).

The third axiom stipulates that dummy museums are excluded from the revenue distribution.

\textbf{Dummy}. For each $(M,N,\pi,E) \in \mathcal{D}$ and each $i \in M$, if $i \in \DM(M,N,\pi,E)$ then
$$
R_i(M,N,\pi,E)=0
$$

The next property is a milder version of previous requirement. It assumes that even dummy museums may be entitled to receive positive revenues. Consider, for example, a small and traditional dummy museum that is important for the city's heritage. In this case, it is natural to desire and apply some degree of social solidarity. In general, from a philosophical and moral perspective, the next axiom states that a society should have some obligations to their dummy individuals, and the straightforward exclusion is, at least, arguable. More precisely, \emph{order preservation with dummies} says that a dummy museum may receive a positive allocation as long as it is smaller than the allocation of any other non-dummy museum. Therefore, this solidarity property respects the priority of the non-dummy museums in receiving part of the total revenue.

\textbf{Order preservation with dummies}. For each $(M,N,\pi,E) \in \mathcal{D}$ and each $i \in M$, if $i \in \DM(M,N,\pi,E)$ then
$$
R_i(M,N,\pi,E) \leq R_j(M,N,\pi,E) \quad \forall j \in M \setminus \DM(M,N,\pi,E)
$$

It is worth noting that order preservation with dummies weakens the dummy in two aspects. Firstly, order preservation with dummies does not exclude the possibility of not assigning anything to dummy museums, it simply does not force this to happen. And secondly, dummy implies that all dummy museums obtain the same (zero) revenue, a restriction that is not imposed in order preservation with dummies.

Although solidarity is a natural requirement, it must be bounded. In this sense, the next axiom strengthens order preservation with dummies. It imposes an upper limit for the award of a dummy museum. More precisely, for any $\tau \in [0,1]$, \emph{$\tau$-order preservation with dummies} says that the allocation of any dummy museum is, at most, the $\tau$-th part of the allocation of any other non-dummy museum.

\textbf{$\tau$-order preservation with dummies}. For each $(M,N,\pi,E) \in \mathcal{D}$ and each $i \in M$, if $i \in \DM(M,N,\pi,E)$ then
$$
R_i(M,N,\pi,E) \leq \tau R_j(M,N,\pi,E) \quad \forall j \in M \setminus \DM(M,N,\pi,E)
$$

Notice that the parameter $\tau$ modulates the degree of solidarity. If $\tau=1$, then $\tau$-order preservation with dummies coincides with order preservation with dummies. If $\tau=0$, then it is equivalent to dummy.

The next property states that the distribution of revenue does not depend on the names of the pass holders.

\textbf{Pass holder anonymity}. For each $(M,N,\pi,E) \in \mathcal{D}$, each permutation of $N$, $\sigma:N \longrightarrow N$, and each $i \in N$,
$$
R_{i}(M,\sigma(N),\pi,\sigma(E)) = R_{\sigma(i)}(M,N,\pi,E)
$$
where $\sigma(N) = \{\sigma(a) : a \in N\}$, and $\sigma(E)$ is the matrix whose rows are reordered according to the permutation $\sigma$.

The property below stipulates that when the set of museums, the set of pass holders, and the pass price are kept fixed, the allocation of a dummy museum is independent of the distribution of the visits.

\textbf{Independence of visits distribution}. For each $(M,N,\pi,E), (M,N,\pi,E') \in \mathcal{D}$, $E \neq E'$, and each $i \in \DM(M,N,\pi,E) \cap \DM(M,N,\pi,E')$
$$
R_i(M,N,\pi,E) = R_i(M,N,\pi,E').
$$


\section{Results}

\subsection{Reduced domain $\overline{\mathcal{D}}$}

\cite{Bergantinos15} showed that there exists a unique rule on $\overline{\mathcal{D}}$ that satisfies equal treatment of equals, revenue additivity, and dummy.

\begin{theorem}\label{GJ-thm1}
	On $\overline{\mathcal{D}}$, a rule satisfies equal treatment of equals, revenue additivity, and dummy if and only if it is the Shapley rule.
\end{theorem}

Our next result states that, if we relax dummy to order preservation with dummies, more rules emerge. In particular, we characterize the convex combinations of the Shapley and the uniform rules. Additionally, the parameter of the combination is not a number but a function that may vary across pass holders, and which is dependent on the museums a pass holder visits.

\begin{theorem}\label{MPP-thm1}
	On $\overline{\mathcal{D}}$, a rule satisfies equal treatment of equals, revenue additivity, and order preservation with dummies if and only if
	$$
	R(M,N,\pi,E) = \sum_{a \in N} \left[ \beta_a(M_a)R^{U} (M,\{a\},\pi,E^{(a)}) + (1-\beta_a(M_a))R^{Sh} (M,\{a\},\pi,E^{(a)}) \right]
	$$
	for some mappings $\beta_a: 2^M \longrightarrow [0,1]$, $a \in N$, where $M_a=\{i \in M: E_{ai}=1\}$ and $E^{(a)}$ is the entrance matrix with just the row corresponding to the pass holder $a\in N$.
	\begin{proof}
		Let $R$ be a rule defined as in the statement. We start by showing that this rule satisfies the three properties.
		\begin{itemize}
			\item Equal treatment of equals. If $E_{ai} = E_{aj}, \forall a \in N$, then it is straightforward that $R_i \left(M,N,\pi, E \right) = R_j \left(M,N,\pi, E \right)$.
			\item Order preservation with dummies. Let $\{i,j\} \subseteq M$ such that $i \in \DM\left(M,N,\pi, E \right)$ and $j \notin \DM\left(M,N,\pi, E \right)$. By definition of the uniform and Shapley rules we have that
			$$
			R_i(M,N,\pi,E) = \sum_{a \in N} \beta_a(M_a) \frac{\pi}{m}
			$$
			and
			$$
			R_j(M,N,\pi,E) = \sum_{a \in N} \beta_a(M_a) \frac{\pi}{m} + \sum_{a \in N} (1-\beta_a(M_a)) \frac{\pi}{|M(a)|}
			$$
			And then $R_i(M,N,\pi,E) \leq R_j(M,N,\pi,E)$.
			\item Revenue additivity. This property follows from the rule's additive structure.
		\end{itemize}
		Now, we prove the converse. Let $R$ be a rule that fulfills \emph{equal treatment of equals}, \emph{revenue additivity}, and \emph{order preservation with dummies}.

		Let $(M,N,\pi,E) \in \overline{\mathcal{D}}$, where $N$ contains only one pass holder, $N=\{a\}$. Due to \emph{equal treatment of equals}, there exist $x_a(M_a), y_a(M_a) \in \mathbb{R}_+$ such that
		$$
		R_i(M,N,\pi,E) =
		\begin{cases}
		x_a(M_a) & \text{if } E_{ai} = 0 \\
		y_a(M_a) & \text{if } E_{ai} = 1
		\end{cases}
		$$
		Given $M_a=\{i \in M: E_{ai}=1\}$, in application of \emph{order preservation with dummies}, there also exists $\alpha_a(M_a) \in [0,1]$ such that $x(M_a)=\alpha_a(M_a) y(M_a)$. Therefore,
		$$
		\sum_{i \in M} R_i(M,N,\pi,E) = \pi \quad \equiv \quad (m-e^a)x_a(M_a) + e^ay_a(M_a)= \pi \quad \equiv \quad y_a(M_a)= \frac{\pi}{\alpha_a(M_a)(m-e^a)+e^a}
		$$
		And hence,
		$$
		R_i(M,N,\pi,E) =
		\begin{cases}
		\dfrac{\alpha_a(M_a) \pi}{\alpha_a(M_a)(m-e^a)+e^a} & \text{if } E_{ai} = 0 \\[0.3cm]
		\dfrac{\pi}{\alpha_a(M_a)(m-e^a)+e^a} & \text{if } E_{ai} = 1
		\end{cases}
		$$
		Let us define $\beta_a(M_a)=\frac{m\alpha_a(M_a)}{\alpha_a(M_a)(m-e^a)+e^a}$. Notice that, since $\alpha_a(M_a) \in [0,1]$, we also have that $\beta_a(M_a) \in [0,1]$. Then,
		$$
		R(M,N,\pi,E) =  \beta_a(M_a)R^{U} (M,N,\pi,E) + (1-\beta_a(M_a))R^{Sh} (M,N,\pi,E)
		$$
		Indeed, if $i \in M$ is such that $E_{ai}=0$,
		\begin{align*}
			R_i(M,N,\pi,E) &= \beta_a(M_a)R^{U} (M,N,\pi,E) + (1-\beta_a(M_a))R^{Sh} (M,N,\pi,E) \\
			&= \beta_a(M_a) \frac{\pi}{m} \\
			&= \frac{\alpha_a(M_a) \pi}{\alpha_a(M_a)(m-e^a)+e^a}
		\end{align*}
		If $i \in M$ is such that $E_{ai}=1$,
		\begin{align*}
			R_i(M,N,\pi,E) &= \beta_a(M_a)R^{U} (M,N,\pi,E) + (1-\beta_a(M_a))R^{Sh} (M,N,\pi,E) \\
			&= \beta_a(M_a) \frac{\pi}{m} + (1-\beta_a(M_a)) \frac{\pi}{e^a} \\
			&= \frac{\pi}{\alpha_a(M_a)(m-e^a)+e^a} + \left( 1-\frac{m\alpha_a(M_a)}{\alpha_a(M_a)(m-e^a)+e^a}  \right)\frac{\pi}{e^a} \\
			&= \frac{\pi}{\alpha_a(M_a)(m-e^a)+e^a} \left(\alpha_a(M_a) + \frac{e^a-\alpha_a(M_a)e^a}{e^a}  \right) \\
			&= \frac{\pi}{\alpha_a(M_a)(m-e^a)+e^a}
		\end{align*}
		Now, let $(M,N,\pi,E) \in \overline{\mathcal{D}}$ without any restriction on $N$. By \emph{revenue additivity}, it follows that
		$$
		R_i \left(M,N,\pi, E \right) = \sum_{a \in N} R_i \left(M,\{a\},\pi, E^{(a)} \right),
		$$
		where each $E^{(a)}$ is the entrance matrix with just the row corresponding to the pass holder $a\in N$. Since we have already derived the expression of each $R_i \left(M,\{a\},\pi, E^{(a)} \right)$, the proof is complete.
	\end{proof}
\end{theorem}

\begin{remark}\label{MPP-remark1}
	The axioms in Theorem \ref{MPP-thm1} are independent.
	\begin{itemize}
		\item Suppose we are on the enlarged domain, $\mathcal{D}$. Let $R^1$ be the rule that, for each pass holder, assigns the price she has paid to the museum with the lowest label among those she has visited. If the pass holder has not visited any museum, then the rule divides the pass price uniformly. That is, for each $(M,N,\pi,E) \in \mathcal{D}$ and each $i \in M$,
		$$
		R^1_i(M,N,\pi,E)=\sum_{a=1}^n z_{ai} \pi,
		$$
		where
		$$
		z_{ai}=
		\begin{cases}
		1 & \text{if } e^a>0 \text{ and } E_{aj}=0 \; \forall j<i\\
		\frac{1}{m} & \text{if } e^a=0 \\
		0 & \text{otherwise}
		\end{cases}
		$$
		This rule $R^P$, restricted to $\overline{\mathcal{D}}$, satisfies revenue additivity and order preservation with dummies. However, it violates equal treatment of equals.
		\item The proportional rule $R^P$, restricted to $\overline{\mathcal{D}}$, satisfies equal treatment of equals and order preservation with dummies. However, it violates revenue additivity.

		\item Suppose we are on the enlarged domain, $\mathcal{D}$. Let $R^2$ be the rule that, for each pass holder, uniformly assigns the price she has paid among the museums she has not visited. In the case that the pass holder has visited every museum, then the rule equally divides the pass price among all of them. That is, for each $(M,N,\pi,E) \in \mathcal{D}$ and each $i \in M$,
		$$
		R^2_i(M,N,\pi,E)=\sum_{a=1}^n z_{ai} \pi,
		$$
		where
		$$
		z_{ai}=
		\begin{cases}
		\frac{1}{m} & \text{if } e^a=m \text{ or } e^a=0 \\
		\frac{1}{m-e^a} & \text{if } 0<e^a<m \text{ and } E_{ai}=0 \\
		0 & \text{otherwise}
		\end{cases}
		$$
		This rule, restricted to $\overline{\mathcal{D}}$, satisfies revenue additivity and equal treatment of equals. However, it violates order preservation with dummies.
	\end{itemize}
\end{remark}

If, in addition to the axioms in the previous result, we also require the rule to satisfy pass holder anonymity and independence of visits distribution, then we must conclude that such a rule is a convex combination of the uniform and Shapley rules where, in this case, the parameter is fixed and does not depend on other elements of the problem.

\begin{theorem}\label{MPP-thm2}
	On $\overline{\mathcal{D}}$, a rule satisfies equal treatment of equals, revenue additivity, order preservation with dummies, pass holder anonymity, and independence visits distribution if and only if
	$$
	R = \beta R^{U}  + (1-\beta) R^{Sh}
	$$
	for some $\beta \in [0,1]$.
	\begin{proof}
		Theorem \ref{MPP-thm1} guarantees that any rule in this family satisfies equal treatment of equals, revenue additivity, and order preservation with dummies. It is straightforward to check that it also fulfills the other two properties. Now, we prove the converse. Let $(M,N,\pi,E) \in \overline{\mathcal{D}}$. By Theorem \ref{MPP-thm1}, we know that there exist $n$ mappings, one for each $a \in N$, $\beta_a: 2^M \longrightarrow [0,1]$, such that
		$$
		R(M,N,\pi,E) = \sum_{a \in N} \left[ \beta_a(M_a)R^{U} (M,N,\pi,E) + (1-\beta_a(M_a))R^{Sh} (M,N,\pi,E) \right]
		$$
		Since $R$ satisfies \emph{pass holder anonymity}, $\beta_a(\cdot) = \beta_b(\cdot)=\beta(\cdot)$ for all $\{a,b\} \subset N$. Hence,
		$$
		R(M,N,\pi,E) = \sum_{a \in N} \left[ \beta(M_a)R^{U} (M,N,\pi,E) + (1-\beta(M_a))R^{Sh} (M,N,\pi,E) \right]
		$$
		By \emph{revenue additivity},
		$$
		R \left(M,N,\pi, E \right) = \sum_{a \in N} R \left(M,\{a\},\pi, E^{(a)} \right),
		$$
		where each $E^{(a)}$ is the entrance matrix with just the row corresponding to the pass holder $a\in N$. Now,
		$$
		R \left(M,\{a\},\pi, E^{(a)} \right) = \beta(M_a)R^{U} \left(M,\{a\},\pi, E^{(a)} \right) + (1-\beta(M_a))R^{Sh} \left(M,\{a\},\pi, E^{(a)} \right)
		$$
		\emph{Independence of visits distribution} implies that $\beta(M'_a)=\beta(M_a)=\beta$ for any $M_a,M'_a \subseteq M$. Indeed, consider two entrance matrices $E^{(a)}$ and $E'^{(a)}$ such that  $M_a=\{i \in M: E^{(a)}_{ai}=1\}$ and  $M'_a=\{i \in M: E'^{(a)}_{ai}=1\}$, respectively. Let $i \in \DM\left(M,N,\pi, E^{(a)} \right) \cap \DM\left(M,N,\pi, E'^{(a)} \right)$. By definition of uniform and Shapley rules, we have that
		$$
		R_i \left(M,\{a\},\pi, E^{(a)} \right) = \beta(M_a) \frac{\pi}{m} \quad \text{ and } \quad  R_i \left(M,\{a\},\pi, E'^{(a)} \right) = \beta(M'_a) \frac{\pi}{m}
		$$
		Since \emph{independence of visits distribution} implies that $R_i \left(M,\{a\},\pi, E^{(a)} \right)=R_i \left(M,\{a\},\pi, E'^{(a)} \right)$, it must happen that $\beta(M_a)=\beta(M'_a)$. Therefore,
		\begin{align*}
			R \left(M,N,\pi, E \right) & = \sum_{a \in N} R \left(M,\{a\},\pi, E^{(a)} \right) \\
			&= \sum_{a \in N} \beta R^{U} \left(M,\{a\},\pi, E^{(a)} \right) + (1-\beta)R^{Sh} \left(M,\{a\},\pi, E^{(a)} \right) \\
			&= \beta R^{U} (M,N,\pi,E) + (1-\beta) R^{Sh} (M,N,\pi,E)
		\end{align*}
	\end{proof}
\end{theorem}

\begin{remark}\label{MPP-remark2}
	The axioms in Theorem \ref{MPP-thm2} are independent.
	\begin{itemize}
		\item The rule $R^1$ defined in Remark \ref{MPP-remark1} satisfies revenue additivity, order preservation with dummies, pass holder anonymity, and independence of visits distribution. However, it violates equal treatment of equals.
		\item The proportional rule $R^P$, restricted to $\overline{\mathcal{D}}$, satisfies equal treatment of equals, order preservation with dummies, pass holder anonymity, and independence of visits distribution. However, it violates revenue additivity.
		\item Let $R^3$ be the rule such that, for each $(M,N,\pi,E) \in \overline{\mathcal{D}}$ and each $i \in M$,
		$$
		R^3(M,N,\pi,E) = \sum_{a \in N} \left[ \beta_a R^{U} (M,\{a\},\pi,E^{(a)}) + (1-\beta_a)R^{Sh} (M,\{a\},\pi,E^{(a)}) \right]
		$$
		for some parameters $\beta_1, \ldots, \beta_n \in [0,1]$. This rule satisfies equal treatment of equals, revenue additivity, order preservation with dummies, and independence of visits distribution. However, it violates pass holder anonymity.
		\item Let $R^4$ be the rule such that, for each $(M,N,\pi,E) \in \overline{\mathcal{D}}$ and each $i \in M$,
		$$
		R^4(M,N,\pi,E) = \sum_{a \in N} \left[ \beta (M_a)R^{U} (M,\{a\},\pi,E^{(a)}) + (1-\beta (M_a))R^{Sh} (M,\{a\},\pi,E^{(a)}) \right]
		$$
		for some mapping $\beta: 2^M \longrightarrow [0,1]$. This rule satisfies equal treatment of equals, revenue additivity, order preservation with dummies, and pass holder anonymity. However, it violates independence of visits distribution.
		\item Let $\varepsilon < \frac{1}{m-1}$, the following rule, $R^{\varepsilon}$, satisfies equal treatment of equals, revenue additivity, pass holder anonymity, and independence of visits distribution. However, it violates order preservation with dummies. For each $i \in M$,
		$$
		R_{i}^{\varepsilon}(M,N,\pi,E) =\sum_{a \in N}\left\{(1-E_{ai})\frac{1+\varepsilon}{m} + E_{ai} \frac{m - (m-e^a)(1+\varepsilon)}{m\cdot e^a}\right\}\pi
		$$
	\end{itemize}
\end{remark}

For the next result, we replace order preservation with dummies by $\tau$-order preservation with dummies, which is a stronger requirement. Theorem \ref{MPP-thm3} states that if this is done, then the rule must also be a convex combination of the uniform and Shapley rules, where the parameter $\beta$ is upper bounded by a value that depends both on the degree of social solidarity $\tau$ and the number of pass holders $n$.

\begin{theorem}\label{MPP-thm3}
	On $\overline{\mathcal{D}}$, a rule satisfies equal treatment of equals, revenue additivity, $\tau$-order preservation with dummies, pass holder anonymity, and independence of visits distribution if and only if
	$$
	R = \beta R^{U}  + (1-\beta) R^{Sh}
	$$
	for some $\beta \in \left[0,\frac{\tau}{n+\tau(1-n)} \right]$.
	\begin{proof}
		It is clear that any rule defined as in the statement satisfies equal treatment of equals, revenue additivity, and pass holder anonymity. Let us check the fulfillment of $\tau$-order preservation with dummies and independence of visits distribution. Indeed,
		\begin{itemize}
			\item $\tau$-order preservation with dummies. Let $i \in \DM(M,N,\pi,E)$ and $j \in M \backslash \DM(M,N,\pi,E)$. Since $R^{Sh}_j(M,N,\pi,E) \geq 0$, applying the expressions of the uniform and Shapley rules, we have that
			$$
			R_i(M,n,\pi,E) = \beta \frac{n\pi}{m} \leq \beta \frac{n\pi}{m} + R^{Sh}_j(M,N,\pi,E) = R_j(M,n,\pi,E)
			$$
			\item Independence of visits distribution. Let $(M,N,\pi,E), (M,N,\pi,E') \in \mathcal{D}$ be two problems such that $E \neq E'$. Let $i \in \DM(M,N,\pi,E) \cap \DM(M,N,\pi,E')$. By definition of the uniform and Shapley rules, we have that
			$$
			R_i(M,N,\pi,E) = \beta \frac{n\pi}{m} + (1-\beta) \cdot 0 = R_i(M,N,\pi,E').
			$$
	  \end{itemize}
		Now, let us prove the converse. Let $R$ be a rule that satisfies the properties in the statement. Notice that \emph{$\tau$-order preservation with dummies} implies social security with dummies. By Theorem \ref{MPP-thm2}, we know that there exists $\beta \in [0,1]$ such that, for any $(M,N,\pi,E) \in \overline{\mathcal{D}}$,
		$$
		R(M,N,\pi,E) = \beta R^{U}(M,N,\pi,E)  + (1-\beta) R^{Sh}(M,N,\pi,E)
		$$
		Let $i \in \DM(M,N,\pi,E)$ and $j \notin \DM(M,N,\pi,E)$. After applying \emph{$\tau$-order preservation with dummies}, we have that
		$$
		R_i(M,N,\pi,E) = \beta \frac{n\pi}{m} \leq \tau \left[ \beta \frac{n\pi}{m}   + (1-\beta) \sum_{a=1}^n \frac{E_{aj}}{e^a} \pi \right] = R_j(M,N,\pi,E)
		$$
		Therefore, $\beta \in [0,1]$ must be such that
		$$
		\beta \leq \frac{m\tau \varepsilon}{n-n\tau+m\tau \varepsilon}, \quad \text{where} \quad \varepsilon = \sum_{a=1}^n \frac{E_{aj}}{e^a}
		$$
		Since the previous inequality must hold for any problem, then $\beta \leq f(\varepsilon^*)$, where $f(\varepsilon) = \frac{m\tau \varepsilon}{n-n\tau+m\tau \varepsilon}$ and $\varepsilon^*=\arg \min_{\varepsilon} f(\varepsilon)$. Given that $f$ is increasing in $\varepsilon$, it is minimized when $\varepsilon$ takes its minimum value. This occurs when there exists a unique $b \in N$ such that $E_{bj}=1$, $E_{aj}=0$ for any other $a \in N \setminus \{b\}$ and $e^b=m$. In such a case, $\varepsilon^*=\frac{1}{m}$. Therefore,
		$$
		\beta \leq \frac{m\tau \varepsilon^*}{n-n\tau+m\tau \varepsilon^*} \quad \Leftrightarrow \quad \beta \leq \frac{\tau}{n+\tau(1-n)}
		$$
	\end{proof}
\end{theorem}

\begin{remark}\label{MPP-remark3}
	The axioms in Theorem \ref{MPP-thm3} are independent.
	\begin{itemize}
		\item The rule $R^1$ defined in Remark \ref{MPP-remark1} satisfies revenue additivity, $\tau$-order preservation with dummies, pass holder anonymity, and independence of visits distribution. However, it violates equal treatment of equals.
		\item The proportional rule $R^P$, restricted to $\overline{\mathcal{D}}$, satisfies \emph{equal treatment of equals}, $\tau$-\emph{order preservation with dummies}, \emph{pass holder anonymity}, and \emph{independence of visits distribution}:However, it violates \emph{revenue additivity}.
		\item Let $R^{3,\tau}$ be the rule such that, for each $(M,N,\pi,E) \in \overline{\mathcal{D}}$ and each $i \in M$,
		$$
		R^{3,\tau}(M,N,\pi,E) = \sum_{a \in N} \left[ \beta_a R^{U} (M,\{a\},\pi,E^{(a)}) + (1-\beta_a)R^{Sh} (M,\{a\},\pi,E^{(a)}) \right]
		$$
		for some parameters $\beta_1, \ldots, \beta_n \in \left[0,\frac{\tau}{n+\tau(1-n)} \right]$. This rule satisfies equal treatment of equals, revenue additivity, $\tau$-order preservation with dummies, and independence of visits distribution. However, it violates pass holder anonymity.
		\item Let $R^{4,\tau}$ be the rule such that, for each $(M,N,\pi,E) \in \overline{\mathcal{D}}$ and each $i \in M$,
		$$
		R^{4,\tau}(M,N,\pi,E) = \sum_{a \in N} \left[ \beta (M_a)R^{U} (M,\{a\},\pi,E^{(a)}) + (1-\beta (M_a))R^{Sh} (M,\{a\},\pi,E^{(a)}) \right]
		$$
		for some mapping $\beta: 2^M \longrightarrow \left[0,\frac{\tau}{n+\tau(1-n)} \right]$. This rule satisfies equal treatment of equals, revenue additivity, $\tau$-order preservation with dummies, and pass holder anonymity. However, it violates independence of visits distribution.
		\item The rule $R^{\varepsilon}$ in Remark \ref{MPP-remark2} satisfies \emph{equal treatment of equals}, \emph{revenue additivity}, \emph{pass holder anonymity}, and \emph{independence of visits distribution}. However, it violates $\tau$-\emph{order preservation with dummies}.
	\end{itemize}
\end{remark}

The previous section mentioned that $\tau$-order preservation with dummies coincides with order preservation with dummies when $\tau=1$, as well as with dummy when $\tau=0$. Accordingly, if we impose $\tau=1$ on the previous result, then $\beta$ is not restricted ($\beta \in [0,1]$) and Theorem \ref{MPP-thm3} becomes Theorem \ref{MPP-thm2}. Similarly, if $\tau=0$, then $\beta=0$ and Theorem \ref{MPP-thm3} becomes Theorem \ref{GJ-thm1}.


\subsection{Enlarged domain $\mathcal{D}$}

For the reduced domain Theorem 1 states that only the Shapley rule satisfies equal treatment, revenue additivity, and dummy. However, the next example shows that there is no rule on the enlarged domain that fulfills dummy.

\begin{example}
Let us consider the museum pass problem with museums $M=\{1,2\}$, pass holders $N=\{1,2\}$, pass price $\pi$, and an entrance matrix given by:
$$
E=\left(
\begin{array}{cc}
0 & 0 \\
0 & 0 \\
\end{array}
\right)
$$
In this case both museums are dummy, and therefore $R_1(M,N,\pi,E)=R_2(M,N,\pi,E)=0$. However, this contradicts the definition of rule, which requires that $R_1(M,N,\pi,E)+R_2(M,N,\pi,E)=\pi$.
\end{example}

Consequently, Theorem \ref{GJ-thm1} cannot be replicated on $\mathcal{D}$. Theorem \ref{MPP-thm1}, on the other hand, can be adapted. The next result characterizes the family of rules that fulfills equal treatment of equals, revenue additivity, and order preservation with dummies on $\mathcal{D}$. Such a family is pretty similar to that obtained in Theorem \ref{MPP-thm1}, except that we replace the Shapley rule by the equal attribution rule.\footnote{We omit the proof because it is similar to the proof for Theorem \ref{MPP-thm1}.}

\begin{theorem}\label{MPP-thm5}
	On $\mathcal{D}$, a rule satisfies equal treatment of equals, revenue additivity, and order preservation with dummies if and only if
	$$
	R(M,N,\pi,E) = \sum_{a \in N} \left[ \beta_a(M_a)R^{U} (M,N,\pi,E) + (1-\beta_a(M_a))R^{EA} (M,N,\pi,E) \right]
	$$
	for some mappings $\beta_a: 2^M \longrightarrow [0,1]$, $a \in N$.
\end{theorem}

\begin{remark}\label{MPP-remark4}
	The axioms in Theorem \ref{MPP-thm5} are independent.
	\begin{itemize}
		\item The rule $R^1$ defined in Remark \ref{MPP-remark1} satisfies, on $\mathcal{D}$, revenue additivity and order preservation with dummies, but it violates equal treatment of equals.
		\item The proportional rule $R^P$ satisfies equal treatment of equals and order preservation with dummies, but it violates revenue additivity.
		\item The rule $R^2$ defined in Remark \ref{MPP-remark1} satisfies, on $\mathcal{D}$, satisfies revenue additivity and equal treatment of equals, but it violates order preservation with dummies.
	\end{itemize}
\end{remark}

In Section \ref{MPP-secrules}, we showed that the Shapley rule cannot be applied to the domain $\mathcal{D}$. We presented several alternatives to extend this rule from $\overline{\mathcal{D}}$ to $\mathcal{D}$. Beyond the mere characterization, Theorem \ref{MPP-thm5} also states that, among all the possible extensions, the equal attribution rule is the unique alternative that is compatible with equal treatment of equals, revenue additivity, and order preservation with dummies.

In view of the previous result, one would expect to have alternative characterizations to Theorems \ref{MPP-thm2} and \ref{MPP-thm3} on $\mathcal{D}$ by simply replacing the Shapley rule by the equal attribution rule. Interestingly, however, this is not the case. The next result states that, on the enlarged domain, the uniform rule is the unique rule that satisfies equal treatment of equal, revenue additivity, and independence of visits distribution. Notice that, in comparison with Theorem \ref{MPP-thm2}, order preservation with dummies and pass holder anonymity are not required for the characterization.

\begin{theorem}\label{MPP-thm6}
On $\mathcal{D}$, a rule satisfies equal treatment of equals, revenue additivity, and independence of visits distribution if and only if it is the uniform rule.
\begin{proof}
	It is straightforward to check that the uniform rule satisfies the properties in the statement. We focus on the interesting implication. Let $(M, \{a\}, \pi, 0^{(a)})$ be a problem with just one pass holder who has not visited any museum. By \emph{equal treatment of equals}, $R_i(M, \{a\}, \pi, 0^{(a)})=\frac{\pi}{m}$ for each $i \in M$. Let $(M, \{a\}, \pi, 1^{(a)})$ be a problem with just one pass holder who has visited all museums. By \emph{equal treatment of equals}, $R_i(M, \{a\}, \pi, 1^{(a)})=\frac{\pi}{m}$ for each $i \in M$. Now, let $(M, \{a\}, \pi, E^{(a)})$ be a problem with just one pass holder such that $M_{a} \neq \phi$ and $M \backslash M_a \neq \phi$, where $M_a=\{i \in M: E^{(a)}_{ai}=1\}$. Let $i \in M \backslash M_a$, by \emph{independence of visits distribution},
	$$
	R_{i}\left(M, \{a\}, \pi, E^{(a)}\right)=R_{i}(M, \{a\}, \pi, 0^{(a)})=\frac{\pi}{m}
	$$
	Let $i \in M_a$, by \emph{equal treatment of equals},
	$$
	R_{i}\left(M, \{a\}, \pi, E^{(a)}\right) = \frac{\pi-(m-|M_a|)\frac{\pi}{m}}{|M_a|} = \frac{\pi}{m}
	$$
	Thus, $R_{i}\left(M, a, \pi, E^{(a)}\right)=\frac{\pi}{m}$ for each $i \in M$. Then,
	$$
	R\left(M, \{a\}, \pi, E^{(a)}\right)=R^{U}\left(M, \{a\}, \pi, E^{(a)}\right)
	$$
	Now, let $(M,N,\pi,E) \in \mathcal{D}$ without any restriction on $N$. By \emph{revenue additivity}, it follows that, for each $i \in M$,
	$$
	R_i \left(M,N,\pi, E \right) = \sum_{a \in N} R_i \left(M,\{a\},\pi, E^{(a)} \right) =  \sum_{a \in N} \frac{\pi}{m} = \frac{n\pi}{m} = R^U_i \left(M,N,\pi, E \right)
	$$
\end{proof}
\end{theorem}

\begin{remark}\label{MPP-remark5}
	The axioms in Theorem \ref{MPP-thm6} are independent.
	\begin{itemize}
		\item Let $R^5$ be the rule such that, for each pass holder, assigns the price she has paid to the museum with the lowest label. This rule satisfies revenue additivity and independence of visits distribution, but it violates equal treatment of equals.
		\item The proportional rule $R^P$ satisfies equal treatment of equals and independence of visits distribution, but it violates revenue additivity.
		\item The equal attribution rule satisfies equal treatment of equals and revenue additivity, but it violates independence of visits distribution.
	\end{itemize}
\end{remark}

Even though there are many rules that satisfy $\tau$-order preservation with dummies and independence of visits distribution on $\overline{\mathcal{D}}$, the combination of both is too demanding on $\mathcal{D}$. The following proposition states that, on $\mathcal{D}$, there is no rule that fulfills those two requirements.

\begin{proposition}\label{MPP-thm7}
On $\mathcal{D}$, there is no rule that satisfies $\tau$-order preservation with dummies and independence of visits distribution when $\tau <1$.
\begin{proof}
Let us fixed the set of museums $M=\{1,2\}$, the set of pass holders $N=\{1,2\}$, and the pass price $\pi=\frac{1}{2}$. Let $R$ be a rule that fulfills $\tau$-order preservation with dummies and independence of visits distribution. Consider the allocation selected by the rule for the following three problems
	$$
	(x_1,x_2) = R \left( M,N,\pi, \left( \begin{array}{cc} 0 & 0 \\ 0 & 0 \end{array} \right) \right),
	$$
	$$
	(y_1,y_2) = R \left( M,N,\pi, \left( \begin{array}{cc} 1 & 0 \\ 1 & 0 \end{array} \right) \right),
	$$
	and
	$$
	(z_1,z_2) = R \left( M,N,\pi, \left( \begin{array}{cc} 0 & 1 \\ 0 & 1 \end{array} \right) \right)
	$$
	\emph{Independence of visits distribution} implies that $y_2=x_2$ and $z_1=x_1$, and then $y_1+x_2=1$ and $x_1+z_2=1$. By \emph{$\tau$-order preservation with dummies} $x_2 \leq \tau y_1$, which implies that $x_1 \leq \frac{\tau}{1+\tau}$. Similarly, $x_2 \leq \frac{\tau}{1+\tau}$. However, if $\tau <1$ then $x_1+x_2=\frac{2\tau}{1+\tau}<1$, which contradicts the definition of rule.
\end{proof}
\end{proposition}

The immediate consequence is that the characterization in Theorem \ref{MPP-thm3} cannot be extended to $\mathcal{D}$. We must remember that the goal of $\tau$-order preservation with dummies is to limit the extent of such a social solidarity from the non-dummies. Proposition \ref{MPP-thm7} shows that the choice of the domain, $\overline{\mathcal{D}}$ or $\mathcal{D}$, has a significant impact. On $\mathcal{D}$, if we try to modulate the social solidarity, then the family of possible rules in Theorem \ref{MPP-thm6} shrinks to the empty set. This, however, was not the case for the reduced domain $\overline{\mathcal{D}}$, on which the family is restricted but not so drastically.

\section{Conclusions}\label{conc}

In this paper, we axiomatically analyzed the problem of sharing the revenues obtained from selling museum passes. We enlarged the domain of problems to encompass cases where one or several pass holders do not visit any museum. We discovered that on this enlarged domain, the characterization of the Shapley rule by \cite{Bergantinos15} does not hold. In fact, it turns into an incompatibility among equal treatment of equals, dummy, and revenue additivity.

As an alternative to dummy, we proposed order preservation with dummies, which admits a strictly positive allotment for the dummy museums as long as the allotment does not exceed the amount obtained by any non-dummy museum. This new axiom translates to this setting an idea of social fairness: even dummy individuals should have some minimal rights. However, it is also natural to impose some limitations to that solidarity, which is what $\tau$-order preservation with dummies does. It places an upper bound on the allocation a dummy museum can receive relative to a non-dummy museum. Both on the reduced and enlarged domains, we characterized the family of rules that satisfies equal treatment of equals, order preservation with dummies, and revenue additivity. These families are convex combinations of the Shapley and uniform rules. We find out that restricting the social solidarity by requiring $\tau$-order preservation with dummies instead of just order preservation with dummies has different implications depending on the considered domain. On the reduced domain this axiom is compatible with the rest of the requirements. On the contrary, they are incompatible in the enlarged domain. This finding shows that the choice of the domain is not a minor issue.

It is possible to isolate the Shapley rule from the rest of the family in Theorem \ref{MPP-thm2}, despite the fact that we do not do this explicitly in the paper. \cite{Beal16} introduced the property of \emph{independence of external visitors}, which says that the allocation of a museum should not be affected by the arrival of a new pass holder who does not visit such a museum. If, in addition to the axioms in Theorem \ref{MPP-thm2}, we also require the rule to satisfy independence of external visitors, then the Shapley rule is the only one, on $\overline{\mathcal{D}}$, that fulfills all the properties. However, not all those axioms are independent of each other. Thus, equal treatment of equals, revenue additivity, and independence of external visitors together characterize the Shapley rule.

In the context of TU-games, other papers have proposed different alternatives to dummy (or \emph{null player}, using the terminology of cooperative games). That is the case in \cite{Brink07}, \cite{Casajus14}, and \cite{Beal16}, among others. These works propose different conditions for a player to be null, but in all those cases, the null player receives nothing. In contrast, order preservation with dummies alters the consequence of being null. For TU-games, \cite{Joosten96} introduced the \emph{egalitarian Shapley value} whose expression is equivalent to the family of rules characterized in Theorem \ref{MPP-thm2}. \cite{Casajus13} characterize the egalitarian Shapley value by means of \emph{efficiency}, \emph{additivity}, \emph{desirability}, and \emph{null player in a productive environment}. In Theorem \ref{MPP-thm2} we use a different collection of properties. In our model, any rule satisfies the principles underlying efficiency and null player in a productive environment. Additivity differs significantly from revenue additivity. order preservation with dummies is milder than desirability. And, in addition, equal treatment of equals, pass holder anonymity, and independence of visits distribution are necessary for our result. \cite{Brink11} also characterize, for TU-games, the egalitarian Shapley value. The properties these authors propose (such as \emph{consistency}, \emph{desirability}, \emph{linearity}, and \emph{weak covariance}) differ from those in Theorem \ref{MPP-thm2}.

Finally, we should acknowledge there are several extensions of the model that are not addressed in this work. For example, there may be other elements to be considered for the allocation method, such as the number of exclusive visitors or the ticket price of each museum. Information is another issue that we do not study in this work. We have assumed that we know the set of museums that each pass holder has visited. However, it may be the case that we only know the number of pass holders that has visited each museum but we ignore each visitor's identity. Additionally, our work proposes to reconsider the fairness principle behind the dummy requirement, in this and other models. Many societies face the problem of allocating scarce resources among their members. Different political ideologies (from communism to liberalism, for example) address this problem from opposed perspectives, which result in very disimilar assignment criteria. We believe that the key element to explain most of the disparities among those political ideologies is how they treat null individuals.

\newpage
\textbf{Compliance with Ethical Standards}

Funding: Ricardo Mart\'inez acknowledges the R\&D\&I project grant PID2020-114309GB-I00 funded by MCIN AEI/10.13039/501100011033 and by "ERDF A way of making Europe/EU", and he also acknowledges financial support from Junta de Andaluc\'ia under projects FEDER UGR-A-SEJ-14-UGR20 and Grupos PAIDI SEJ660. Joaqu\'in S\'anchez-Soriano acknowledges the R\&D\&I project grant PGC2018-097965-B-I00 funded by "MCIN/AEI/10.13039/501100011033" and by "ERDF A way of making Europe/EU", and he also acknowledges financial support from the Generalitat Valenciana under project PROMETEO/2021/063.

Conflict of Interest: Both authors declares that they have no conflict of interest.

Ethical approval: This article does not contain any studies with human participants or animals performed by any of the authors.

\newpage

\end{document}